\documentclass[12pt]{article}

\def\tf#1#2{{\textstyle{#1 \over #2}}}

\def\sqr#1#2{{\vcenter{\vbox{\hrule height.#2pt
         \hbox{\vrule width.#2pt height#1pt \kern#1pt
            \vrule width.#2pt}
         \hrule height.#2pt}}}}
\def\square{\mathop{\mathchoice\sqr56\sqr56\sqr{3.75}4\sqr34\,}\nolimits}
\begin{document}
\baselineskip=15.5pt
\pagestyle{plain}
\setcounter{page}{1}

%Title page

\begin{titlepage}
\bigskip
\rightline{PUPT-2107}
\rightline{hep-th/0401041}
\bigskip\bigskip\bigskip
\centerline{\Large \bf Integrable Open Spin Chains}
\medskip
\centerline{\Large \bf in Defect Conformal Field Theory}
\bigskip\bigskip\bigskip

\centerline{\large Oliver DeWolfe$^a$ and Nelia Mann$^b$}
\bigskip
\centerline{\em $^a$ Department of Physics, Princeton University,} \centerline{\em Princeton, NJ 08544} \centerline{\em odewolfe@princeton.edu}
\smallskip
\centerline{\em $^b$ Department of Physics, University of California,} \centerline{\em Santa Barbara, CA 93106} \centerline{\em nelia@physics.ucsb.edu}
\bigskip
\bigskip
\bigskip\bigskip

%ABSTRACT

\begin{abstract}
We demonstrate that the one-loop dilatation generator for the scalar sector of a certain perturbation of ${\cal N} =4$ Super Yang-Mills with fundamentals is the Hamiltonian of an integrable spin chain with open boundary conditions.  The theory is a supersymmetric defect conformal field theory (dCFT) with the fundamentals in hypermultiplets confined to a codimension one defect.  We obtain a $\mathcal{K}$-matrix satisfying a suitably generalized form of the boundary Yang-Baxter equation, study the Bethe ansatz equations, and demonstrate how Dirichlet and Neumann boundary conditions arise in field theory, and match to existing results in the plane wave limit.
\medskip
\noindent
\end{abstract}
\end{titlepage}

\section{Introduction}

${\cal N}=4$ Super Yang-Mills theory in four dimensions has a
number of fascinating properties.  It is exactly
superconformal. In the case of an $SU(N)$ gauge group (to which we
restrict ourselves here) it is believed to be precisely dual to
Type IIB string theory in an $AdS_5 \times S^5$ background with
$N$ units of five-form flux \cite{Maldacena}, \cite{GKP}, \cite{Witten}, \cite{MAGOO}.  This duality
becomes more tractable to study in the limit of excitations with large $R$-charge,
where the gravity dual geometry reduces to a maximally
supersymmetric plane wave, as investigated by Berenstein,
Maldacena, and Nastase~\cite{BMN}.  In this regime it is
particularly easy to see how a closed string arises as a
collection of ``bits," identified in the dual as fields in a gauge invariant single-trace operator.

More recently it has emerged that an integrable structure is
present on both sides of this duality.  The planar one-loop matrix
of anomalous dimensions for gauge invariant local operators has
been shown to be the Hamiltonian of an integrable spin chain; the
case of scalar operators was demonstrated by Minahan and
Zarembo~\cite{MZ}, and their arguments were generalized to all
gauge invariant local operators by Beisert and
Staudacher~\cite{BS}.  Furthermore, on the string theory side the
worldsheet Green-Schwarz theory was shown by Bena, Polchinski, and
Roiban~\cite{BPR} to possess an infinite number of nonlocal conserved quantities.  The two manifestations of integrability were related by Dolan,
Nappi, and Witten~\cite{DNW}; more recent work on integrability and
$AdS$/CFT includes \cite{Belitsky}-\cite{Kim}.

One natural question to ask is whether the integrability of ${\cal
N}=4$ SYM is present in other field theories.  Systems closely
related to ${\cal N}=4$ are the most likely to share in this
special property; for other results on this question see \cite{WW}, \cite{Roiban}.  A particularly interesting feature to introduce
is fundamental matter, associated with open strings in the gravity
dual.

One model with these properties is the defect conformal field theory (dCFT) formulated in~\cite{DFO}; for more work on defect field theories see \cite{EGK}-\cite{Bab}.  This
system couples  the four-dimensional fields and dynamics of ${\cal N}=4$ SYM  to $M$ new three-dimensional hypermultiplets localized on a flat codimension one hypersurface, or ``defect."  These hypermultiplets transform in the fundamental representation of the three-dimensional part of the ${\cal N}=4$ gauge field, and break half the supersymmetry.  There is only one free parameter, the bulk gauge coupling, and interestingly, the theory is exactly
superconformal~\cite{DFO},~\cite{EGK}.  For $M \ll N$, the gravity
dual is IIB string theory on $AdS_5 \times S^5$, with the new
ingredient of $M$ D5-branes wrapped on $AdS_4 \times
S^2$ \cite{KR1}, \cite{KR2}.\footnote{For $M \sim N$ the backreaction of the D5-branes
must be taken into account, and the supergravity solution is
unknown; but see~\cite{Fayyaz}, \cite{CH}.}  Open string modes on the
D5-brane are dual to gauge-invariant operators localized on the
defect, which have the form of a number of bulk adjoint fields
between a pair of fundamental defect fields; the defect fields
thus play the role of the ends of the open string.

In this paper, we consider the question of whether this dCFT is
integrable, as ${\cal N}=4$ SYM is.  We study the sector of
gauge-invariant, local defect operators composed of Lorentz
singlet fields, generalizing the approach of \cite{MZ}.  After
calculating the one-loop anomalous dimension matrix in the planar
limit for this class of operators, we search for an integrable
system with this operator as the Hamiltonian.  This comes down to
finding integrable boundary conditions for the $SO(6)$ spin chain
of~\cite{MZ}, which can be codified in a matrix $\mathcal{K}$ that
solves the boundary Yang-Baxter equation, suitably generalized to
incorporate boundary flavor degrees of freedom.\footnote{For work on classification of integrable boundary conditions without this generalization, see \cite{Doikou1}, \cite{Doikou2}.}  We find such a
$\mathcal{K}$-matrix and demonstrate that the associated
Hamiltonian indeed matches the one-loop dilatation generator of
the dCFT, demonstrating integrability for our sector of operators.
We conjecture that this integrability persists to the complete set
of one-loop gauge invariant operators.

Having obtained the integrable structure, one may use the Bethe
ansatz to diagonalize the anomalous dimension matrix.  Studying
single excitations on the open chain and their reflections off the
boundary, we demonstrate that they obey either Dirichlet or
Neumann boundary conditions, matching the possible
excitations of the dual open string.  Furthermore, the
periodicities and phase shifts conspire so that the the
propagation of left- and right-movers on the chain is equivalent
to that of only left-movers on an effective periodic closed chain of twice
the length---the familiar so-called doubling trick of open string
theory.  The integrability of the system thus provides a window
through which we may see more directly how open strings are realized
within the dual dCFT.  We also compare our results to the results
of Lee and Park in the plane-wave limit~\cite{LP}, and find
agreement.

In section~\ref{FieldTheorySec} we review the defect conformal
field theory and calculate the planar one-loop matrix of anomalous
dimensions.  We solve the boundary Yang-Baxter equation to
identify the integrable structure in section~\ref{YangBaxterSec},
and formulate the Bethe ansatz and consider single excitations in
section~\ref{BetheSec}.  In section~\ref{ConclusionSec} we conclude.  Field theory conventions are listed in an appendix.

{\em Note added.}  As this paper was being finalized there appeared work by Chen, Wang, and Wu~\cite{CWW} that obtains apparently similar results for a different deformation of ${\cal N}=4$ SYM with fundamentals.
Interesting earlier work on open spin chains in gauge theories with matter appears in \cite{Braun}-\cite{Derk}.

\section{Field Theory Computations}
\label{FieldTheorySec}

\subsection{Review of Defect Conformal Field Theory}

The defect theory we study~\cite{DFO} is a descendant of ${\cal
N}=4$ Super-Yang-Mills theory with $SU(N)$ gauge group.
The ${\cal N}=4$ theory contains a gauge field $A_\mu$, and
adjoint Majorana spinors $\lambda^\alpha$ and real scalars $X^i$
in the ${\bf 4}$ and ${\bf 6}$ respectively of the $SO(6)$ R-symmetry.
The dCFT introduces $M$ additional 3D hypermultiplets propagating
on the hypersurface $x_3 = 0$.  The new fields break the total
symmetry group from $PSU(2,2|4)$ down to $OSp(4|4)$, where the 4D conformal group
$SO(4,2)$ is reduced to the 3D conformal group $SO(3,2)$, the R-symmetry is broken
$SO(6)_R \rightarrow SO(3)_H \times SO(3)_V$, and the
supersymmetry is cut in half.  The defect fields are $M$ complex
scalars $q^m$ transforming in the $({\bf 2}, {\bf 1})$ of $SO(3)_H
\times SO(3)_V$ and $M$ complex 2-component fermions $\Psi^a$
transforming in the $({\bf 1}, {\bf 2})$.  Conventions and the action for ${\cal N}=4$ SYM can be found in the appendix.

The bulk fields are arranged with respect to the preserved 3D
symmetry group into a vector multiplet $\{ A_k, P_+
\lambda^\alpha, X_V^A, D_3 X_H^I \}$ and a hypermultiplet $\{ A_3,
P_- \lambda^\alpha, X_H^I, D_3 X_V^A \}$, where $k=0,1,2$, $A =
1,2,3$, and $I = 4,5,6$.  $P_+ \lambda$ and $P_- \lambda$ with
$P_\pm \equiv (1 \pm \gamma^5 \gamma^3)/2$ indicate the splitting
of a 4D Majorana spinor into two 3D Majorana spinors. The $X_H$
and $X_V$ scalars transform as $({\bf 3}, {\bf 1})$ and $({\bf 1},
{\bf 3})$ of $SO(3)_H \times SO(3)_V$, while the $\lambda^\alpha$
are arranged into a $({\bf 2}, {\bf 2})$ field $\lambda_{am}$.

The defect fields couple directly only to the bulk vector
multiplet.  The total action for the theory is that of ${\cal
N}=4$ SYM together with
\begin{eqnarray}
\label{3DAction}
S_3 &=& S_{kin} + S_{yuk} + S_{pot} \,, \\
S_{kin} &=& \tf{1}{g^2} \int d^3x \left( (D^k q^m)^\dagger D_k q^m
- i \bar\Psi^a \rho^k D_k \Psi^a \right) \,, \\
S_{yuk} &=& \tf{1}{g^2} \int d^3x \left(i \bar{\Psi}^a P_+ \lambda_{a m} q^m - i \bar{q}^m \bar{\lambda}_{m a} P_+ \Psi^a   + \bar\Psi^a \sigma^A_{ab} X_V^A \Psi^a\right)  \,, \\
\label{Spot}
S_{pot} &=&
 \tf{1}{g^2} \int d^3x \left( \bar{q}^m X_V^A X_V^A q^m \, +
i \epsilon_{IJK} \bar{q}^m \sigma^I_{mn} X_H^J X_H^K q^n \right) \\
&& + \tf{1}{g^2} \int d^3x \left(\bar{q}^m \sigma^I_{mn} (D_3 \, X_H^I)\, q^n +  \tf14  \, \delta(0) \, {\rm Tr} \,  (\bar{q}^m \sigma^I_{mn} q^n)^2  \right)
\,, \nonumber
\end{eqnarray}
with $D_k q = \partial_k q - i A_k q$ and similarly for $\Psi$.
The total theory has only one coupling $g$, which is exactly
marginal~\cite{EGK}.  We have written (\ref{3DAction}) -
(\ref{Spot}) for $M=1$, which is sufficient for our purposes; for
$M>1$, dual to multiple D5-branes, flavor indices appear on $q$
and $\Psi$ in an obvious way.

\subsection{Anomalous Dimensions for Scalar Operators}

We are interested in local, gauge-invariant defect operators
constructed out of Lorentz scalar fields that are ``single-trace"
in the sense that one line of color runs through the entire
operator. These have the form
\begin{eqnarray}
\label{DefectOps}
{\cal O}  = \psi_{m,j_1, \ldots j_L, n} \; \bar{q}_m X^{j_1} \ldots X^{j_L} q^n \,,
\end{eqnarray}
with the obvious sum over color indices implied. We can think of
each operator as an open spin chain consisting of vector spaces
$W_{\bar{\alpha}} \times V_1 \times \ldots \times V_L \times
W_\beta$, with $V_i$ in the ${\bf 6}$ of $SO(6)$, $\beta$ in the
${\bf 2}$ of $SO(3)_H$, and $\bar{\alpha}$ in the ${\bf
\bar{2}}$.\footnote{The ${\bf \bar{2}}$ is equivalent to the ${\bf
2}$, but we write it this way to avoid worrying about epsilon
symbols.}

We are interested in the one-loop planar contributions to the
matrix of anomalous dimensions for these operators. We calculate
this by evaluating the correlation function
\begin{eqnarray}
\Gamma \equiv \langle \bar{q}_{n'} (z_\beta) X^{i_L}(z_L) \ldots X^{i_1}(z_1) q^{m'}(z_\alpha) \; {\cal O}(0) \rangle \,.
\end{eqnarray}
Only nearest neighbor fields along the chain will interact to this
order, reducing the contributions to effective correlators of two
fields with a two-site piece of the operator ${\cal O}$.

First consider the interaction between two bulk scalars $X^{i_a}$
and $X^{i_b}$.  Interactions that exist
purely in the ${\cal N}=4$ theory were calculated for the case of the
closed spin chain \cite{MZ} (see also \cite{BMN}, \cite{GMR}, \cite{KPSS}, \cite{G7}) and include the nontrivial interactions of gauge boson exchange and the 4-point $X$ interaction, as well as
contributions of the one-loop wavefunction renormalization of each
$X$ field.  In principle, additional interactions involving the
defect fields can contribute as well.  However, one can quickly
convince oneself by looking at the action (\ref{3DAction}) that
there are no nontrivial defect interactions between 4 $X$'s at
lowest order.  Furthermore, although diagrams with defect fields
contributing to $\langle X X \rangle$ do exist, they are forbidden
on symmetry grounds from leading to a divergent piece in the $X$
self-energy~\cite{DFO}.

Consequently, the contribution of two $X$ fields to the matrix of
anomalous dimension is unchanged from the pure ${\cal N}=4$ case.
As a result, all the interactions in $\Gamma$ not involving a
defect field match the contributions in the case of the
closed chain. (This also implies that the anomalous dimension matrix for
closed chains of $X$'s in the defect theory takes exactly the
same form as in ${\cal N}=4$ SYM.)  The defect interactions are localized
to the ends of the chain and will have the interpretation of
integrable boundary conditions added to an already integrable
system.

\subsection{Differential Regularization and Bulk Interactions}

Although they are well known \cite{MZ}, we include the
calculation of the $XX$ interactions to establish our conventions.
We perform our computations in position space using differential
regularization~\cite{FJL}.  The presence of the defect favors a
position space regulator, and differential regularization is
particularly convenient for extracting anomalous dimensions.

The essential philosophy of differential regularization is to
replace amplitudes containing position-space distributions too
singular to have a Fourier transform by regularized distributions
differing only at the singular point; this replacement is equivalent to 
adding local counterterms.  The regularized amplitudes have a well-defined
Fourier transform but depend on an arbitrary mass scale;
differentiating with respect to this scale produces the usual
Callan-Symanzik equations for the regulated amplitudes.  As we are 
interested only in anomalous dimensions, we keep only logarithmic
divergences, and discard quadratic divergences.

Consider first the amputated $X$ self-energy $\Gamma_X \equiv
\langle X^i(z_1) X^j(z_2) \rangle$.  There are three one-loop
diagrams---fermion loop, gauge boson loop, and scalar tadpole; the
last is a purely quadratic divergence and hence irrelevant for us.
The fermion  and gauge boson diagrams give
\begin{eqnarray}
\Gamma_X^{(1)} = g^2 N \delta^{ij} ( 4 - 2) \partial^\mu \Delta_{12} \partial_\mu \Delta_{12} = g^2 N \delta^{ij} {8 \over (4 \pi^2)^2} {1 \over z_{12}^6} \,,
\end{eqnarray}
where we discarded a quadratic divergence going like $\Delta_{12}
\square \Delta_{12} \sim \delta(z_{12})/z_{12}^2$.  Regularizing
according to the formulae in the Appendix, we find
\begin{eqnarray}
M {\partial \over \partial M} \Gamma_X^{(1)} = \delta^{ij} {g^2 N \over 8 \pi^2} \square \delta(z_{12}) \,.
\end{eqnarray}
Since $\beta(g) = 0$ for our dCFT, the Callan-Symanzik equation is
particularly simple:
\begin{eqnarray}
\left(M {\partial \over \partial M} + 2 \gamma_X(g) \right) \Gamma_X = 0 \,.
\end{eqnarray}
Using the amputated zeroth-order result $\Gamma_X^{(0)} = -
(1/2) \delta^{AB} \square \delta(z_{12})$, we find
\begin{eqnarray}
\label{gammaX}
\gamma_X(g) = {g^2 N \over 8 \pi^2} \,.
\end{eqnarray}
To find the contribution of a pair of $X$ fields to the anomalous
dimension of the defect operator, we compute the correlator of two
X fields with the appropriate part of ${\cal O}(0)$:
\begin{eqnarray}
\Gamma_4 \equiv \langle X^{i_1}(z_1) X^{i_2}(z_2) \, X^{j_1} X^{j_2} (0) \rangle \,.
\end{eqnarray}
Amputating the external legs associated with $X^{i_1}$ and
$X^{i_2}$, regularizing, and keeping only the divergent parts, we
arrive at
\begin{eqnarray}
M {\partial \over \partial M} \Gamma_4^{(1)} = {g^2 N^2 \over 32 \pi^2} \delta(z_1) \delta(z_2) \left( ({\cal T}_A)^{i_1 i_2}_{j_1 j_2} + ({\cal T}_X)^{i_1 i_2}_{j_1 j_2} + ({\cal T}_\Sigma)^{i_1 i_2}_{j_1 j_2} \right)\,,
\end{eqnarray}
where the tensor structures coming from the gauge boson exchange,
4-point X interaction, and self-energy between $X^{i_1}$ and
$X^{j_1}$ are
\begin{eqnarray}
({\cal T}_A)^{i_1 i_2}_{j_1 j_2} =  \delta^{i_1}_{j_1} \delta^{i_2}_{j_2} \,, \quad
({\cal T}_X)^{i_1 i_2}_{j_1 j_2} = 2 \delta^{i_1}_{j_2} \delta^{i_2}_{j_1} - \delta^{i_1 i_2} \delta_{j_1 j_2} - \delta^{i_1}_{j_1} \delta^{i_2}_{j_2} \,, \quad
({\cal T}_\Sigma)^{i_1 i_2}_{j_1 j_2} = - 4 \delta^{i_1}_{j_1} \delta^{i_2}_{j_2} \,.
\end{eqnarray}
Only one of the two possible self-energies is included, so that the other may be included in the interaction of the next pair in the chain.  Correspondingly, we include only one factor $\gamma_X$ from the Callan-Symanzik equation for the total correlator $\Gamma$:
\begin{eqnarray}
\left( M {\partial \over \partial M} + \gamma_X + \gamma_{\cal O} \right) \Gamma_4 = 0 \,.
\end{eqnarray}
Using this equation, the zero-loop expression $\Gamma_4^{(0)} = N \delta^{i_1}_{j_1} \delta^{i_2}_{j_2} \delta(z_1) \delta(z_2)$,
and the result (\ref{gammaX}) for $\gamma_X$, we obtain
\begin{eqnarray}
\label{BulkAnomDim}
(\gamma_{\cal O})^{i_1 i_2}_{j_1 j_2} = {g^2 N \over 16 \pi^2} \left\{
2 \delta^{i_1}_{j_1} \delta^{i_2}_{j_2} - 2 \delta^{i_1}_{j_2} \delta^{i_2}_{j_1} + \delta^{i_1 i_2} \delta_{j_1 j_2}\right\} \,.
\end{eqnarray}
Summed over all nearest neighbor pairs, (\ref{BulkAnomDim}) forms
the complete one-loop planar dilatation operator for a closed
chain~\cite{MZ}.

\subsection{Defect Interactions}

We turn now to the defect interactions.  One necessary ingredient
is the $q$ self-energy.  Defining $\Gamma_q \equiv \langle q^m
(y_1) \bar{q}_n(y_2) \rangle$, we find 4 diagrams at one-loop
order: a $\Psi/\lambda$ fermion loop, a $\partial_3 X_H$ loop, a
gauge boson loop, and a $q$ tadpole; the last diagram does not
contribute to the logarithmic divergence.  The contributions to
the amputated self-energy are
\begin{eqnarray}
\Gamma_{q, \Psi/\lambda}^{(1)} &=& 2 g^2 N \delta^m_n   {\rm Tr} \, (S_{12} \hat{s}_{12})  \,, \\
\Gamma_{q, \partial_3 X_H}^{(1)} &=& - { 3 g^2 N \over 2} \delta^m_n {\cal D}_{12}  \left( \partial_x^2 \Delta_{12} \right) \! \Big|_{x=0}  \,, \\
\Gamma_{q, A_\mu}^{(1)} &=& - g^2 N \delta^m_n \left( 2 \partial^k {\cal D}_{12} \partial_k \Delta_{12} + 2 \Delta_{12} \nabla^2 {\cal D}_{12} + \tf12 {\cal D}_{12} \nabla^2 \Delta_{12}  \right) \,,
\end{eqnarray}
leading to the total logarithmic divergence:
\begin{eqnarray}
\Gamma_q^{(1)} = {g^2 N \over (4 \pi)(4 \pi^2)} \delta^m_n ( 8 + 3 - 5) {1 \over |y_{12}|^5} \quad \rightarrow \quad M {\partial \over \partial M} \Gamma_q^{(1)} = {g^2 N \over 4 \pi^2} \delta^m_n \nabla^2 \delta(y_{12}) \,,
\end{eqnarray}
after regularizing according to the formulae in the appendix.  The
amputated zeroth-order propagator is $- \delta^m_n \nabla^2
\delta(y_{12})$, and hence the Callan-Symanzik equation implies
\begin{eqnarray}
\left( M {\partial \over \partial M} + 2 \gamma_q(g) \right) \Gamma_q = 0 \quad \rightarrow \quad \gamma_q = {g^2 N \over 8 \pi^2} \,.
\end{eqnarray}
Next we consider the interaction at one end of the chain.  We
calculate
\begin{eqnarray}
\Gamma_{\partial} \equiv \langle X^i (z_1) \, q^{m'} (y_2) \, \bar{q}_m X^j (0) \rangle \,.
\end{eqnarray}
A universal, flavor-blind interaction is gauge boson exchange, which leads to
\begin{eqnarray}
g^2 N^2 \delta^{m'}_m \delta^{ij} \left( \Delta_{12} \partial_k \Delta_1 \partial^k {\cal D}_2 - \tf14 \Delta_1 {\cal D}_2 \nabla^2 \Delta_{12} + \tf12 \Delta_1 \partial^k {\cal D}_2 \partial_k \Delta_{12} - \tf12 {\cal D}_2 \partial^k \Delta_1 \partial_k \Delta_{12} \right) \,.
\end{eqnarray}
If one pulls out derivatives to obtain divergence-free total
derivatives plus a term with all derivatives on ${\cal D}_2$ which contains the divergence, you get
\begin{eqnarray}
\Gamma_{\partial, A_\mu}^{(1)} = {\rm total \; derivatives} \ +
{g^2 N^2 \over 4 (4 \pi^2)^2} \delta^{m'}_m \delta^{ij}
\delta(y_2) {1 \over z_1^4} \quad \rightarrow \quad M {\partial
\over \partial M} \Gamma_{\partial, A_\mu}^{(1)} = {g^2 N^2 \over
32 \pi^2} \delta^{m'}_m \delta^{ij} \delta(z_1) \delta(y_2)  \,.
\end{eqnarray}
There is another diagram with a scalar interaction $\bar{q} q X
X$, which is different depending on whether the two $X$ scalars
are $X_V$ or $X_H$.  For $X^i = \delta^i_A X_V^A, X^j = \delta^j_B
X_V^B$, the $q$ and $X$ scalars are charged under different
$SO(3)$s and the flavor interaction must be trivial.  We find
\begin{eqnarray}
\Gamma_{\partial, X_V}^{(1)} = - {g^2 N^2 \over 4} \delta^{m'}_m \delta^{AB} \delta(z_{12}) \Delta_1 {\cal D}_1 \quad \rightarrow \quad M {\partial \over \partial M} \Gamma_{\partial, X_V}^{(1)} = - {g^2 N^2 \over 16 \pi^2} \delta^{m'}_m \delta^{AB} \delta(z_1) \delta(y_2) \,.
\end{eqnarray}
In the case $X^i = \delta^i_I X_H^I, X^j = \delta^j_K X_H^J$ the flavor interaction is nontrivial:
\begin{eqnarray}
\Gamma_{\partial, X_H}^{(1)} = - {g^2 N^2 \over 4} \delta(z_{12}) (i \epsilon_{IJK} \sigma^K_{m' m}) {\cal D}_2 \, \Delta_2 \quad \rightarrow \quad M {\partial \over \partial M} \Gamma_{\partial, X_H}^{(1)} = - {g^2 N^2 \over 16 \pi^2} \delta(z_1) \delta(y_2) (i \epsilon_{IJK} \sigma^K_{m' m}) \,.
\end{eqnarray}
The Callan-Symanzik equation for the complete correlator $\Gamma$ is
\begin{eqnarray}
\label{TotalCS}
\left( M {\partial \over \partial M} + L  \gamma_X + 2 \gamma_q  + \gamma_{\cal O} \right) \Gamma = 0 \,,
\end{eqnarray}
where $\Gamma$ and $\gamma_{\cal O}$ are understood to be
matrix-valued, while the other terms are proportional to the
identity.  The matrix $\gamma_{\cal O}$ can be separated into the
parts acting in the bulk of the chain on two $X$'s, and the
boundary pieces.  There are $L-1$ pairs of neighboring $X$'s, each
of which contributes to the matrix as (\ref{BulkAnomDim}). Notice,
however, that  this does not take into account one factor of
$\gamma_X$ in (\ref{TotalCS}), as well as one contribution of the
$X$ self-energy to $\Gamma$; this is because the open chain has no
$X^{i_L} X^{i_1}$ interaction.  Because these contributions are
proportional to the identity in flavor-space, we can insert them
anywhere;  we find the convenient choice to be to include half the
contribution of each with each boundary term.

The boundary interaction at the left end of the chain is then
determined by
\begin{eqnarray}
\left( M {\partial \over \partial M} + {\gamma_X \over 2} + \gamma_q+ \gamma_{\cal O} \right) \Gamma_q + {1 \over 2} M {\partial \over \partial M} \Gamma^{(1)}_{4, \Sigma} = 0 \,.
\end{eqnarray}
We obtain
\begin{eqnarray}
(\gamma_{\cal O})^{m' I_1 \ldots}_{m J_1 \ldots} = {g^2 N \over 16 \pi^2} \left( 2 \delta^{m'}_m \delta^{I_1}_{J_1} + 2 i \epsilon_{I_1 J_1 K} \sigma^K_{m' m} \right) \,, \quad \quad
(\gamma_{\cal O})^{m' A_1 \ldots}_{m B_1 \ldots} = {g^2 N \over 16 \pi^2} \left( 4 \delta^{m'}_m \delta^{A_1}_{B_1} \right) \,,
\end{eqnarray}
and analogously for the other end of the chain,
\begin{eqnarray}
(\gamma_{\cal O})^{\ldots I_L n'}_{\ldots J_L n} = {g^2 N \over 16 \pi^2} \left( 2 \delta^{n'}_n \delta^{I_L}_{J_L} - 2 i \epsilon_{I_L J_L K} \sigma^K_{n n'} \right) \,, \quad \quad
(\gamma_{\cal O})^{\ldots A_L n' }_{\ldots B_L n} = {g^2 N \over 16 \pi^2} \left( 4 \delta^{n'}_n \delta^{A_L}_{B_L} \right) \,,
\end{eqnarray}
where delta-functions in the unwritten indices are understood.

In summary, we find the matrix of anomalous dimensions to be
\begin{eqnarray}
\label{Gamma}
\gamma_{\cal O} &=& {g^2 N \over 16 \pi^2} \left( \sum_{a=1}^{L-1} h_{a a+1} + (2 I_{\bar\alpha 1} + 2\bar{S}_{\bar\alpha 1}) + (2I_{L \beta} + 2S_{L \beta})  \right) \,, \\
h_{a a+1} &=& K_{a a+1} + 2 I_{a a+1} - 2 P_{a a+1} \,,
\end{eqnarray}
where subscripts on operators indicate the vector spaces they act on, and
\begin{eqnarray}
\nonumber
K^{j_a j_b}_{i_a i_b} = \delta^{j_a j_b} \delta_{i_a i_b} \,, \quad P^{j_a j_b}_{i_a i_b} &=& \delta^{j_a}_{i_b} \delta^{j_b}_{i_a} \,, \quad I^{j_a j_b}_{i_a i_b} = \delta^{j_a}_{i_a} \delta^{j_b}_{i_b} \,, \\
\label{KPI}
S^{I m}_{J n} =   - i \epsilon_{IJK} \sigma^K_{nm} \,, \quad S^{A m}_{B n} &=& \delta^m_n \delta^A_B \,, \quad S^{I m}_{B n} = S^{A m}_{J n} = 0 \,, \\
\bar{S}^{\bar{m} I}_{\bar{n} J} =   i \epsilon_{IJK} \sigma^K_{\bar{m}\bar{n}} \,, \quad \bar{S}^{\bar{m} A}_{\bar{n} B} &=& \delta^{\bar{m}}_{\bar{n}} \delta^A_B \,, \quad \bar{S}^{\bar{m} I}_{\bar{n} B} = \bar{S}^{\bar{m} A}_{\bar{n} J} = 0 \,.
\nonumber
\end{eqnarray}
A check of our computation is that the anomalous dimensions should
vanish for protected operators.  In~\cite{DFO}, the chiral
primaries of the dCFT were determined to be
\begin{eqnarray}
\label{CPO}
\bar{q}^m \sigma^{(I_1}_{mn} X_H^{I_2} X_H^{I_3} \ldots X_H^{I_{L+1})} q^n \,,
\end{eqnarray}
where parentheses denote total symmetrization and tracelessness
among all the triplets of $SO(3)$.  One may indeed verify that
these operators are annihilated by $\gamma_{\cal O}$ as given in
(\ref{Gamma}); in fact each $h_{a, a+1}$, as well as $2
I_{\bar\alpha 1} + 2\bar{S}_{\bar\alpha 1}$  and  $2I_{L \beta} +
2S_{L \beta}$, separately gives zero.

\section{The Boundary Yang-Baxter Equation}
\label{YangBaxterSec}

Having calculated the one-loop planar dilatation generator for the
class of defect operators (\ref{DefectOps}), we want to learn
whether it is the Hamiltonian of an integrable open spin chain.

It was shown by Minahan and Zarembo in~\cite{MZ} that the
operators composed of closed chains of $X$ fields possess this
kind of integrable structure.  This was done by identifying an
$\mathcal{R}$-matrix, in this case the $SO(6)$-invariant
\begin{eqnarray}
\label{SO6R}
\mathcal{R}_{12}(u) = \frac{1}{2}\left[u(u - 2)I_{12} - (u - 2)P_{12} + uK_{12}\right] \,,
\end{eqnarray}
where $1$ and $2$ label the two vector spaces acted on by the
operator, and $I$, $P$, and $K$ are as defined in (\ref{KPI}). These
satisfy the Yang-Baxter equation
\begin{eqnarray}
\label{YangBaxter}
\mathcal{R}_{12}(u)\mathcal{R}_{13}(u + v)\mathcal{R}_{23}(v) = \mathcal{R}_{23}(v)\mathcal{R}_{13}(u + v)\mathcal{R}_{12}(u) \,.
\end{eqnarray}
If we define the transfer matrix as the trace of the monodromy matrix,
\begin{eqnarray}
t(u) = {\rm Tr}_a\, T_a(u) \equiv {\rm Tr}_a\; \mathcal{R}_{a1}(u)\mathcal{R}_{a2}\cdots \mathcal{R}_{aL}(u) \,,
\end{eqnarray}
where $a$ is an auxiliary space and the dependence of $t(u)$ and
$T_a(u)$ on the vector spaces $1,2,\ldots L$ comprising the chain
is understood, the Yang-Baxter equation guarantees that the
transfer matrices commute for arbitrary arguments:
\begin{eqnarray}
[t(u), t(v)] = 0 \,.
\end{eqnarray}
The expansion of the transfer matrix in powers of $u$ then
generates an infinite number of conserved quantities, starting
with the momentum and Hamiltonian.  The identification of the
one-loop dilatation operator for closed chains as the Hamiltonian
associated with (\ref{SO6R}) demonstrated the integrability for
that system.

The analogous method for introducing integrable boundary conditions was formulated in \cite{Skly}; for a pedagogical summary see \cite{DN} or section 3.5 of \cite{book}.  In addition to an integrable $\mathcal{R}$-matrix determining the dynamics for the bulk of the
chain, one introduces operators $\mathcal{K}^{\pm}_a(u)$ acting on
either end of the spin chain.  These satisfy the boundary
Yang-Baxter equations, or BYBs,
\begin{eqnarray}
\label{BoundaryYangBaxter}
\mathcal{R}_{12}(u-v)\mathcal{K}^{-}_{1}(u)\mathcal{R}_{12}(u+v)\mathcal{K}^{-}_{2}(v) &=& \mathcal{K}^{-}_{2}(v)\mathcal{R}_{12}(u+v)\mathcal{K}^{-}_{1}(u)\mathcal{R}_{12}(u-v) \, \\
\mathcal{R}_{12}(v-u)\mathcal{K}^{+
t_1}_{1}(u)\mathcal{R}_{12}(-u-v-2i\gamma)\mathcal{K}^{+
t_2}_{2}(v) &=& \mathcal{K}^{+
t_2}_{2}(v)\mathcal{R}_{12}(-u-v-2i\gamma)\mathcal{K}^{+
t_1}_{1}(u)\mathcal{R}_{12}(v-u)\,. \nonumber
\end{eqnarray}
Here, $t_i$ refers to taking the transpose on the $i$th vector
space, and $\gamma$ is a parameter characteristic of a given
$\mathcal{R}$-matrix such that
\begin{eqnarray}
\mathcal{R}^{t_1}_{12}(u)\mathcal{R}^{t_1}_{12}(-u - 2i\gamma) =
\lambda(u) \,
\end{eqnarray}
for some scalar function $\lambda(u)$ (for our case it is easy to
see that $\gamma = 2i$). We can then form the open chain transfer
matrices
\begin{eqnarray}
\hat{t}(u) = {\rm Tr}_a\,
\mathcal{K}^{+}_{a}(u)T_{a}(u)\mathcal{K}^{-}_{a}(u)T^{-1}_{a}(-u)
\,.
\end{eqnarray}
If (\ref{YangBaxter}) and (\ref{BoundaryYangBaxter}) are satisfied,
these commute,
\begin{eqnarray}
\label{TransferCommute}
[\hat{t}(u), \hat{t}(v)] = 0 \,,
\end{eqnarray}
and thus generate the family of commuting charges characteristic
of an integrable system.

We must generalize this formalism slightly, from matrices
$\mathcal{K}^{-}_1$ and $\mathcal{K}^{+}_L$ that act only on the
last ``ordinary" vector space in the chain, to matrices
$\mathcal{K}^{+}_{\bar{\alpha} 1}$ and $\mathcal{K}^{-}_{L \beta}$
that also act on the degrees of freedom $W_{\bar\alpha}$ and
$W_\beta$ living on the boundary.  In this case the resultant
$\hat{t}(u)$ acts on the vector space $W_{\bar{\alpha}} \times V_1
\times \cdots \times V_L \times W_{\beta}$.

The structure of integrability is largely unaffected by this
modification; the extra indices ``go along for the ride'' in the
BYB, and (\ref{TransferCommute}) still
holds.  Given the quantum numbers of $W_{\bar{\beta}}$ and $V_L$,
the matrix $\mathcal{K}^{-}(u)$ must take the form
\begin{eqnarray}
(\mathcal{K}^-)^{\bar{m} I}_{\bar{n} J}(u) &=& f(u)\, \delta^{\bar{m}}_{\bar{n}}\delta^{I}_{J} + g(u) \, \epsilon_{IJK} \, \sigma^{K}_{\bar{m}\bar{n}} \,, \\
(\mathcal{K}^-)^{\bar{m} A}_{\bar{n} B}(u) &=& h(u) \, \delta^{\bar{m}}_{\bar{n}}\delta^{A}_{B} \,, \\
(\mathcal{K}^-)^{\bar{m} I}_{\bar{n} B}(u) &=& ({\mathcal K}^-)^{\bar{m} A}_{\bar{n} J}(u) = 0 \,,
\end{eqnarray}
where $\bar{m}, \bar{n}$ are indices in $W_{\bar{\beta}}$, $I, J$
are indices in the ${\bf 3}$ of $SO(3)_{H}$, and $A, B$ are
indices in the ${\bf 3}$ of $SO(3)_{V}$.  Furthermore, if we have
a $\mathcal{K}^{-}$ that satisfies the first BYB equation (\ref{BoundaryYangBaxter}), then
$(\mathcal{K}^+_1)^{t_1}(u) = \mathcal{K}^-_1(2 - u)$ satisfies
the second BYB equation;  however, this process
will give matrices acting on a spin chain whose boundaries are
both in the anti-fundamental of $SU(2)$.  We can easily translate
one of the boundary actions to act on the associated fundamental,
and we will do so later.

Thus what we need to do is find what functions $f(u)$, $g(u)$, and
$h(u)$, if any, satisfy the first BYB.  Writing out the BYB with
indices, we want
\begin{eqnarray}
\mathcal{R}^{i_1 i_2}_{j_1 j_2}(u-v)(\mathcal{K}^{-})^{\bar{m} j_1}_{\bar{n} k_1}(u)\mathcal{R}^{k_1 j_2}_{\ell_1 k_2}(u+v)(\mathcal{K}^{-})^{\bar{n} k_2}_{\bar{p} \ell_2}(v) = (\mathcal{K}^{-})^{\bar{m} i_2}_{\bar{n} j_2}(v)\mathcal{R}^{i_1 j_2}_{j_1 k_2}(u+v)(\mathcal{K}^{-})^{\bar{n} j_1}_{\bar{p} k_1}(u)\mathcal{R}^{k_1 k_2}_{\ell_1 \ell_2}(u-v).
\end{eqnarray}
There are six independent indices in this equation: $i_1, i_2,
\ell_1, \ell_2$ and $\bar{m}, \bar{p}$.  The first four can each
take a value in either $SO(3)_{H}$ or $SO(3)_{V}$, allowing for a
total of 16 equation types that need to be satisfied.

It is not hard to see that both sides will vanish whenever three
of the indices $i_1, i_2, \ell_1, \ell_2$ belong to one of the
$SO(3)$'s and the fourth to the other $SO(3)$.  This leaves us
with 8 equations to check.  Out of concern for the reader, we will
give only the equations and their solution, omitting details.  The
case where all indices belong to the $SO(3)_V$ turns out to be
satisfied independently of the form of the $f(u)$, $g(u)$, and
$h(u)$, as do the cases with $i_1, \ell_1 \in SO(3)_H$ or $i_2,
\ell_2 \in SO(3)_H$.

The same two conditions arise from either $\ell_1, \ell_2 \in SO(3)_H$ or  $i_1, i_2 \in SO(3)_H$:
\begin{eqnarray}
\nonumber
(u+v)[(u-v)^2 + 2]&[h(v)f(u) - h(u)f(v)] + (u-v)[(u+v)^2 + 2][h(u)h(v) - f(u)f(v)] \\
 &= 2(u+v+1)(u-v)(u+v-2)g(u)g(v) \,,
\end{eqnarray}
\vskip-.5in
\begin{eqnarray}
\nonumber
(u^2 - v^2)(u-v-2)[h(v)g(u) + h(u)g(v)] + (u+v)(u-v-2)[h(v)g(u) - h(u)g(v)] \\
 - (u^2 - v^2)(u+v-2)[g(u)f(v) - g(v)f(u)] - (u-v)(u+v-2)[g(u)f(v) + g(v)f(u)]  \\
  = - i(u+v+1)(u-v)(u+v-2)g(u)g(v) - 3(u^2 - v^2)g(v)(h(u) + f(u)) \,.
  \nonumber
\end{eqnarray}
The cases $i_1, \ell_2, \in SO(3)_H$ and $i_2, \ell_1 \in SO(3)_H$ give two more equations:
\begin{eqnarray}
(v-u)[f(u)f(v) - h(u)h(v) - 2g(u)g(v)] &=& (v+u)[h(u)f(v) - h(v)f(u)] \,, \\
(v-u)[f(u)g(v) + f(v)g(u) - ig(u)g(v)] &=& (v+u)[h(u)g(v) - h(v)g(u)] \,.
\end{eqnarray}
The case with all indices in $SO(3)_H$, which is the most labor-intensive, gives two final equations:
\begin{eqnarray}
ig(u)g(v)(u - v)[2u + 2v - 1] &=& 2vf(v)g(u) - 2uf(u)g(v) \,,
\end{eqnarray}
\vskip-.5in
\begin{eqnarray}
\nonumber
ig(u)g(v)(u-v)(u+v-2)[2(u+v)(u-v-2) + u + v + 1]   =& \\
2vg(u)f(v)[u^2 - v^2 + 2] - 2g(v)f(u)[u(u^2 - v^2)  - 3(u^2 - v^2) + 2u] -&\\
 3g(v)[f(u) + h(u)](u^2 - v^2) & \,.
\nonumber
\end{eqnarray}
We find that
\begin{eqnarray}
f(u) = 2u^2 - u + 1 \,, \quad \quad
g(u) = 2iu  \,, \quad \quad
h(u) = -2u^2 + u + 1
\end{eqnarray}
satisfy these conditions, up to an overall function of $u$ that we
fix by requiring $\mathcal{K}^{-}_{L \bar{\beta}}(0) =
I_{\bar{\beta} L}$.  (This overall ambiguity corresponds to
reshuffling the basis of the infinite set of mutually commuting
operators inside $\hat{t}(u)$.)  We therefore obtain the result
\begin{eqnarray}
(\mathcal{K}^-)^{\bar{m} I}_{\bar{n} J}(u) &=& (2u^2 - u + 1)\delta^{\bar{m}}_{\bar{n}}\delta^{I}_{J} + 2u \, i  \epsilon_{IJK} \,\sigma^{K}_{\bar{m}\bar{n}} \,, \\
(\mathcal{K}^-)^{\bar{m} A}_{\bar{n} B}(u) &=& (-2u^2 + u + 1)\delta^{\bar{m}}_{\bar{n}}\delta^{A}_{B} \,, \quad \quad
(\mathcal{K}^-)^{\bar{m} I}_{\bar{n} B}(u) = (\mathcal{K}^-)^{\bar{m} A}_{\bar{n} J}(u) = 0 \,,
\nonumber
\end{eqnarray}
and this implies
\begin{eqnarray}
(\mathcal{K}^+)^{\bar{m} I}_{\bar{n} J}(u) &=& (2u^2 - 7u + 7)\delta^{\bar{m}}_{\bar{n}}\delta^{I}_{J} + (2iu-4i) \epsilon_{IJK} \, \sigma^{K}_{\bar{m}\bar{n}} \,, \\
(\mathcal{K}^+)^{\bar{m} A}_{\bar{n} B}(u) &=& (-2u^2 + 7u - 5)\delta^{\bar{m}}_{\bar{n}}\delta^{A}_{B} \,, \quad \quad
(\mathcal{K}^+)^{\bar{m} I}_{\bar{n} B}(u) = (\mathcal{K}^+)^{\bar{m} A}_{\bar{n} J}(u) = 0 \,.
\nonumber
\end{eqnarray}
Remembering that we want to deal with a spin chain having one boundary in the fundamental, we need to determine, given the above $\mathcal{K}^{-}$ acting on the anti-fundamental, a $\tilde{\mathcal{K}}^{-}$ that acts on the fundamental.  For $SU(2)$, if $\psi$ is in the fundamental, then $\sigma^{2}\psi$ is in the anti-fundamental; therefore, we should have $\tilde{\mathcal{K}}^{-} = \sigma^{2}\mathcal{K}^{-}\sigma^{2}$.  Recalling that $\sigma^{2}\sigma^K \sigma^{2} = -(\sigma^K)^{T}$, we obtain
\begin{eqnarray}
(\tilde{\mathcal{K}}^-)^{m I}_{n J}(u) &=& (2u^2 - u + 1)\delta^{m}_{n}\delta^{I}_{J} - 2u \, i  \epsilon_{IJK} \,\sigma^{K}_{n m} \,, \\
(\tilde{\mathcal{K}}^-)^{m A}_{n B}(u) &=& (-2u^2 + u +
1)\delta^{m}_{n}\delta^{A}_{B} \,, \quad \quad
(\tilde{\mathcal{K}}^-)^{m I}_{n B}(u) =
(\tilde{\mathcal{K}}^-)^{m A}_{n J}(u) = 0 \,. \nonumber
\end{eqnarray}
Now, we want to form the transfer matrix
\begin{eqnarray}
\hat{t}(u) = {\rm Tr}_a\left[\mathcal{K}^{+}_{\bar{\alpha} a}(u)\mathcal{R}_{a 1}(u)\cdots \mathcal{R}_{a L}(u)\tilde{\mathcal{K}}^{-}_{a \beta}(u)(\mathcal{R}_{L a}(-u))^{-1}\cdots(\mathcal{R}_{1 a}(-u))^{-1}\right] \,.
\end{eqnarray}
Since (as is straightforward to verify)
\begin{eqnarray}
\mathcal{R}^{-1}_{ba}(-u) = \frac{4}{(4-u^2)(1-u^2)}\mathcal{R}_{ab}(u) \,,
\end{eqnarray}
we find
\begin{eqnarray}
\hat{t}(u) ={\rm  Tr}_{a}\, \left[\frac{4^{L}}{(4-u^2)^L(1-u^2)^L}\mathcal{K}^{+}_{\bar{\alpha} a}(u)\mathcal{R}_{a 1}(u)\cdots \mathcal{R}_{a L}(u)\tilde{\mathcal{K}}^{-}_{a \beta}(u)\mathcal{R}_{a L}(u)\cdots \mathcal{R}_{a 1}(u)\right] \,.
\end{eqnarray}
We are interested in expanding $\hat{t}(u) = \sum_n \hat{t}_n
u^n$, and identifying the Hamiltonian as $\hat{t}_1$. Making use
of the relations
\begin{eqnarray}
\mathcal{R}_{ab}(0) = P_{ab} \,, \quad \quad
\frac{d}{du}\mathcal{R}_{ab}(0) = \frac{1}{2}\left[-2I_{ab} - P_{ab} + K_{ab}\right] \,,
\end{eqnarray}
and the definitions in (\ref{KPI}), we determine that the lowest
term in the expansion of $\hat{t}(u)$ is
\begin{eqnarray}
\label{t0}
\hat{t}_0 \equiv \hat{t}(0) = {\rm Tr}_a\left[\frac{4^{L}}{4^{L}}\mathcal{K}^{+}_{\bar{\alpha} a}(0)P_{a 1}\cdots P_{a L}I_{L \beta}P_{a L}\cdots P_{a 1}\right]
 = {\rm Tr}_a\left[\mathcal{K}^{+}_{\bar{\alpha} a}\right] = 6I \,.
\end{eqnarray}
The next term is slightly more complicated.  Breaking it into pieces, we find that

\begin{eqnarray}
{\rm Tr}_{a}\,\left[\partial_{u}\mathcal{K}^{+}_{\bar{\alpha} a}(0)\mathcal{R}_{a 1}(0)\cdots \mathcal{R}_{a L}(0)\tilde{\mathcal{K}}^{-}_{a \beta}(0)\mathcal{R}_{a L}(0)\cdots \mathcal{R}_{a 1}(0)\right] =\\
  {\rm Tr}_{a}\, \left[\partial_{u}\mathcal{K}^{+}_{\bar{\alpha} a}(0)P_{a 1}\cdots P_{a L}I_{a \beta}P_{a L}\cdots P_{a 1}\right]   = {\rm Tr}_{a}\,\left[\partial_{u}\mathcal{K}^{+}_{\bar{\alpha} a}(0)\right] = 0 \,,
\nonumber
\end{eqnarray}
and
\begin{eqnarray}
\nonumber
{\rm Tr}_{a}\,\left[\mathcal{K}^{+}_{\bar{\alpha} a}(0)\partial_{u}\mathcal{R}_{a 1}(0)\cdots \mathcal{R}_{a L}(0)\tilde{\mathcal{K}}^{-}_{a \beta}(0)\mathcal{R}_{a L}(0)\cdots \mathcal{R}_{a 1}(0)\right]  +\\
\nonumber
{\rm Tr}_{a}\,\left[\mathcal{K}^{+}_{\bar{\alpha} 1}(0)\mathcal{R}_{a 1}(0)\cdots \mathcal{R}_{a L}(0)\tilde{\mathcal{K}}^{-}_{a \beta}(0)\mathcal{R}_{a L}(0)\cdots \partial_{u}\mathcal{R}_{a 1}(0)\right] = \\
\frac{1}{2}{\rm Tr}_{a}\,\left[\mathcal{K}^{+}_{\bar{\alpha} 1}(0)(-2I_{a 1} - P_{a 1} + K_{a 1})P_{a 2}\cdots P_{a L}I_{a \beta}P_{a L}\cdots P_{a 1}\right] + \\
 \nonumber
 \frac{1}{2}{\rm Tr}_{a}\,\left[\mathcal{K}^{+}_{\bar{\alpha} a}(0)P_{a 1}\cdots P_{a L}I_{\beta a}P_{a L}\cdots P_{a 2}(-2I_{a 1} - P_{a 1} + K_{a 1})\right] = \\
{\rm Tr}_{a}\,\left[\mathcal{K}^{+}_{\bar{\alpha} a}(0)(-2I_{a 1} - P_{a 1} + K_{a 1})\right]
 = 12\bar{S}_{\alpha 1} - 13I_{\alpha 1} \,.
\nonumber
\end{eqnarray}
The terms with the derivative acting on $\mathcal{R}_{a \ell}$ for
$\ell \in 2, ..., L$ all behave similarly:
\begin{eqnarray}
\nonumber
{\rm Tr}_{a}\left[\mathcal{K}^{+}_{\bar{\alpha} a}(0)\mathcal{R}_{a 1}(0)\cdots \partial_{u}\mathcal{R}_{a \ell}(0)\cdots \mathcal{R}_{a L}(0)\tilde{\mathcal{K}}^{-}_{a \beta}(0)\mathcal{R}_{a L}(0)\cdots \mathcal{R}_{a 1}(0)\right]  + \\
\nonumber
{\rm Tr}_{a}\left[\mathcal{K}^{+}_{\bar{\alpha} a}(0)\mathcal{R}_{a 1}(0)\cdots \mathcal{R}_{a L}(0)\tilde{\mathcal{K}}^{-}_{a \beta}(0)\mathcal{R}_{a L}(0)\cdots \partial_{u}\mathcal{R}_{a \ell}(0)\cdots \mathcal{R}_{a 1}(0)\right] = \\
 \frac{1}{2} {\rm Tr}_{a}\left[\mathcal{K}^{+}_{\bar{\alpha} a}(0)P_{a 1}\cdots(-2I_{a \ell} - P_{a \ell} + K_{a \ell})\cdots P_{a L}I_{a \beta}P_{a L}\cdots P_{a 1}
\right] + \\
\nonumber
 \frac{1}{2}{\rm Tr}_{a}\left[\mathcal{K}^{+}_{\bar{\alpha} a}(0)P_{a 1}\cdots P_{a L}I_{\beta a}P_{a L}\cdots(-2I_{a \ell} - P_{a \ell} + K_{a \ell})\cdots P_{a_1}\right] = \\
\nonumber
 {\rm Tr}_{a}\left[\mathcal{K}^{+}_{\bar{\alpha} a}(0)P_{a 1}\cdots P_{a \ell-1}\left[-2P_{a \ell} - I_{a \ell} + K_{a \ell}\right]P_{a \ell-1}\cdots P_{a 1}\right] = \\
 {\rm Tr}_{a}\left[\mathcal{K}^{+}_{\bar{\alpha} a}(0)\right](-2P_{\ell-1 \ell} - I_{\ell-1 \ell} + K_{\ell-1 \ell})
 = 6(-2P_{\ell-1 \ell} - I_{\ell-1 \ell} + K_{\ell-1 \ell}) \,.
\nonumber
\end{eqnarray}
Finally, we have a term of the form
\begin{eqnarray}
{\rm Tr}_a [\mathcal{K}^{+}_{\bar{\alpha} a}(0)\mathcal{R}_{a 1}(0)\cdots \mathcal{R}_{a L}(0)\partial_{u}\tilde{\mathcal{K}}^{-}_{a \beta}(0)\mathcal{R}_{a L}(0) \cdots \mathcal{R}_{a L}(0)] \end{eqnarray}
\vskip-.5in
\begin{eqnarray}
\nonumber
 {\rm Tr}_a\left[\mathcal{K}^{+}_{\bar{\alpha} a}(0)P_{a 1}\cdots P_{a L}\partial_{u}\tilde{\mathcal{K}}^{-}_{a \beta}(0)P_{a L}\cdots P_{a 1}\right]
 &=& {\rm Tr}_a\left[\mathcal{K}^{+}_{\bar{\alpha} a}(0)P_{a 1}\cdots\partial_{u}\tilde{\mathcal{K}}^{-}_{L \beta}(0)\cdots P_{a 1}\right] = \\
 {\rm Tr}_a\left[\mathcal{K}^{+}_{\bar{\alpha} a}(0)\right]\partial_{u}\tilde{\mathcal{K}}^{-}_{L \beta}(0)
 &=& 6(-I_{L \beta} + 2S_{L \beta}) \,.
\nonumber
\end{eqnarray}
We combine these expressions to give
\begin{eqnarray}
\hat{t}_1 = -13I_{\bar{\alpha} 1} + 12\bar{S}_{\bar{\alpha} 1} + \Sigma_{\ell = 1}^{L-1}\left(-12P_{\ell, \ell+1} - 6I_{\ell, \ell+1} + 6K_{\ell, \ell+1}\right) - 6I_{\beta L} + 12S_{L \beta} \,,
\end{eqnarray}
and  adding a multiple of the identity to the result,  we find
\begin{eqnarray}
\hat{t}_1 + (25 + 18L)I = 6\left[\Sigma_{\ell = 1}^{L-1} h_{\ell \ell+1} + (2I_{\bar{\alpha} 1} + 2\bar{S}_{\bar{\alpha} 1}) + (2I_{\beta L} + 2S_{L \beta})\right] \,,
\end{eqnarray}
which is indeed proportional to the matrix of anomalous dimension $\gamma_{\mathcal{O}}$ found in (\ref{Gamma}), demonstrating that this operator is in fact a Hamiltonian for an integrable system.

\section{The Bethe Ansatz}
\label{BetheSec}

Having seen that the planar one-loop matrix of anomalous
dimensions (\ref{Gamma}) is an integrable Hamiltonian for the spin
chain, we may use the techniques of the Bethe ansatz to find the
eigenvectors and eigenvalues of this matrix.

To do so one begins with a Bethe reference
state, corresponding to a certain operator, and systematically
creates excitations above this ground state.  These excitations
correspond to impurities in the chain.  The ``Bethe ansatz"  is to
assume that the eigenvectors consist of waves of impurities (or ``spin
waves") propagating along the chain with certain momenta $k_i$; integrability assures that
scattering of these waves factorizes into products of two-body
scattering where the magnitude of each momentum is separately conserved.

The Bethe ansatz requires for consistency one equation for each
excitation.  For the closed chain, these equations are of the
form
\begin{eqnarray}
\label{CBA}
e^{ik_i L} = \prod_{j \neq i} S_{ji}(k_j, k_i),
\end{eqnarray}
where the $S$-matrix $S_{ji}(k_j, k_i)$ is the phase shift
acquired when the $i^{th}$ excitation passes through the $j^{th}$
excitation; the equations (\ref{CBA}) can be thought of as the
requirement that the phase of an excitation as it travels all the
way around the chain and comes back to itself is unity.  In the
special case of non-interaction, $S_{ji} = 1$ and (\ref{CBA})
reduces to the usual quantization of momentum on a circle.

In the case of systems with Lie group symmetry, extra excitations
are introduced that correspond to changing the orientation in
group space of the spin wave.  For example, consider the general
formula given in  \cite{MZ}, eqn.~(4.35):
\begin{eqnarray}
\label{LieBA}
\left(\frac{u_{q,i}+i\vec\alpha_q\cdot \vec w/2}{u_{q,i}-i\vec\alpha_q\cdot \vec w/2}\right)^L = \prod_{j\ne i}^{n_q}\frac{u_{q,i}-u_{q,j}+i\vec\alpha_q\cdot\vec\alpha_q/2}{u_{q,i}-u_{q,j}-i\vec\alpha_q\cdot\vec\alpha_q/2}
\prod_{q'\ne q}\prod_{j}^{n_{q'}}\frac{u_{q,i}-u_{q',j}+i\vec\alpha_q\cdot\vec\alpha_{q'}/2}{u_{q,i}-u_{q',j}-i\vec\alpha_q\cdot\vec\alpha_{q'}/2} \,.
\end{eqnarray}
Here the $u_{q,i}$ are parameters characterizing excitations,
taking the place of the $k_i$; $i$ labels the excitation as
before, while $q$ reflects the fact that the excitation can be
associated to any of the simple roots $\vec\alpha_q$ of the
algebra. $\vec{w}$ is the highest weight vector of the
representation of the group that lives at each site.  For a
fundamental at each site, we will have $\vec{w} = \vec{w}^1$, the
first fundamental weight, which has the inner product with simple
roots $\vec\alpha_q \cdot \vec{w}^1 = \delta_q^1$.

Thus for an excitation $u_{q,i}$ with $q=1$, the left hand side is
\begin{eqnarray}
\left(\frac{u_{1,i}+i/2}{u_{1,i}-i/2}\right)^L,
\end{eqnarray}
which allows us to match the equation to (\ref{CBA}) by using the
relation between $u_{1,i}$ and the quasi-momentum $k_i$
(\cite{MZ}, eqn.~(4.39)):
\begin{eqnarray}
k(u_{1,i})= -i \log { u_{1,i} + i/2 \over u_{1,i} - i/2} \,.
\end{eqnarray}
Note that the momenta and energy depend only on the $u_{1,i}$
excitations.  For the $u_{q,i}$, $q \neq 1$ excitations, the left
hand side of (\ref{LieBA}) becomes trivial, but the right hand
side constrains the possible $u_{q,i}$.  One should think of the
$u_{1,i}$ as equivalent to the quasi-momenta $k_i$, and creating
an actual excitation with energy and momentum, while the other
$u_{q,i}$ do not change the energy or momentum but instead change
the Lie group quantum numbers of the excitation.

The open chain Bethe ansatz equations generalize (\ref{CBA}) in an
intuitive way.  In the open case, instead of making one complete circuit of the
closed chain, picking up phases $S_{ji}$ for each interaction, a given
excitation passes  one way across the chain, reflects off
one boundary, passes back the other way across the chain, and
reflects off the other boundary:
\begin{eqnarray}
e^{2ik_i L} = {\cal B}_1(k_i) {\cal B}_2(k_i) \prod_{j \neq i} \left( S_{ji}(k_j, k_i) S_{ji}(k_j, -k_i) \right) \,.
\end{eqnarray}
Here ${\cal B}_1$ and ${\cal B}_2$ are phases generated by reflecting 
off the boundary, while the $S_{ji}$ are the same
$S$-matrix terms as in the closed chain; the interaction between
the spin waves is ignorant of the boundary conditions.  We get
each $S_{ji}$ twice because we pass both ways along the chain,
with the relative momentum reversed between the two interactions.

So we see that if one already knows about the closed chain, the
$S_{ji}$ are known, and determining the open chain Bethe ansatz
equations comes down to finding ${\cal B}_1$ and ${\cal B}_2$.  In
our case the $S_{ji}$ are determined by (\ref{LieBA}) for group
$SO(6)$, as was studied in \cite{MZ}.  The
reflection coefficients ${\cal B}_1$ and ${\cal B}_2$ involve only
the excitation in question and the boundary; whether there are
other excitations is irrelevant.  They can hence be determined by
considering the case with just a single excitation.  We do this
below.

\subsection{Bethe Reference State}

We use as our Bethe reference state a particular chiral primary of the dCFT.  Defining $Z \equiv X_H^1 + i X_H^2$, we take
\begin{eqnarray}
\label{Betheref}
|0 \rangle_L \equiv \bar{q}_1 Z Z \ldots Z q_2 \,,
\end{eqnarray}
which results from (\ref{CPO}) where for each $a = 1 \ldots L+1$
one has chosen $I_a = +$, in the standard basis where $J_3$ is the
Cartan generator of $SO(3)_H$.  This state has $j_H = L+1$, $j_V
=0 $.

\subsection{Single Impurity Sector---$SO(3)_V$ Fluctuation}

We now consider the case where any single $Z$ field in the chain
is replaced by another $X^i$ such that $j_H = L$.\footnote{Note
that replacing a $Z$ by $\bar{Z}$ actually leads to $j_H = L-1$;
this mixes with two $X^3$ impurities and hence will not be
included in this section.}  There are two ways we can do this: we
can excite one of the $SO(3)_V$ scalars, or we can excite $X_H^3$.

First consider an $SO(3)_V$ excitation; since the Bethe state is
invariant under $SO(3)_V$, any of the three scalars are
equivalent, and we will let $W$ be any of $X_V^4$, $X_V^5$, or
$X_V^6$. We define the state $|W(x)\rangle$ of the spin chain as
follows:
\begin{eqnarray}
| W(x) \rangle \equiv \bar{q}_1 Z Z \ldots W \ldots Z q_2 \,,
\end{eqnarray}
where $W$ has replaced $Z$ at the $x^{th}$ site.  We then make a
standard Bethe ansatz for the energy eigenstates $|W(k) \rangle$:
\begin{eqnarray}
|W(k) \rangle \equiv \sum_{x = 1}^L f(x) |W(x) \rangle \,, \quad \quad
f(x) = A(k) e^{ikx} + \tilde{A}(k) e^{-ikx} \,.
\label{Bethef}
\end{eqnarray}
Unlike the closed chain ansatz, where we need consider waves
propagating only in one direction, this case requires, because of reflection, 
that both directions be present.  Integrability underlies the
assumption that the magnitude $k$ of the momentum will not change,
either in interacting with another excitation or in reflecting off
the boundary.

Now we wish to determine the action of the Hamiltonian  on the
position eigenstates $|W(x)\rangle$; for convenience we work with
$H$ defined as $\gamma_{\cal O} \equiv (g^2 N /16 \pi^2) \, H$.
Consider first the case where $x \neq1, L$; the boundary parts $2
I_{\bar\alpha 1} + R_{\bar\alpha 1}$, $2 I_{L \beta} + 2 S_{L
\beta}$ of $H$ act on the same fields $\bar{q}_1 Z$ and $Z q_2$ as
in the chiral primary, and hence give zero.  The same is true for
$h_{j,j+1}$ acting on a pair $Z Z$.  Hence the only nonzero
contributions to the Hamiltonian are
\begin{eqnarray}
h_{x, x+1} |W(x) \rangle = (K_{x,x+1} + 2 I_{x,x+1} -2 P_{x,x+1} |W(x) \rangle = 0 + 2 |W(x)\rangle - 2 |W(x+1)\rangle \,, \\
h_{x-1, x} |W(x) \rangle = (K_{x-1,x} + 2 I_{x-1,x} -2 P_{x-1,x} |W(x) \rangle = 0 + 2 |W(x)\rangle - 2 |W(x-1)\rangle \,,
\end{eqnarray}
leading to
\begin{eqnarray}
H |W(x) \rangle = 4 |W(x) \rangle - 2 |W(x-1)\rangle - 2 |W(x+1)\rangle \,.
\end{eqnarray}
Now consider the states $|W(1)\rangle$ and $|W(L)\rangle$.
Nonzero contributions are
\begin{eqnarray}
h_{1,2} |W(1) \rangle =  2 |W(1)\rangle - 2 |W(2)\rangle \,, &&
(2 I_{\bar\alpha 1} + 2 R_{\bar\alpha 1}) |W(1) \rangle = 4 |W(1)\rangle \,, \\
h_{L-1,L} |W(1) \rangle = 2 |W(L)\rangle - 2 |W(L-1)\rangle \,,  &&
(2 I_{L \beta} + 2 R_{L \beta}) |W(L) \rangle = 4 |W(L)\rangle \,.
\nonumber
\end{eqnarray}
Thus we have
\begin{eqnarray}
H |W(1)\rangle = 6 |W(1) \rangle - 2 |W(2)\rangle \,, \quad \quad H |W(L) \rangle = 6 |W(L) \rangle - 2 |W(L-1) \rangle \,.
\end{eqnarray}
Now we demand that the states $|W(k)\rangle$ be energy eigenstates:
\begin{eqnarray}
(H - E(k)) |W(k) \rangle = 0 \,,
\end{eqnarray}
implying the relations
\begin{eqnarray}
\label{VRelations}
0 &=& (4 - E(k)) f(x) - 2 f(x+1) - 2 f(x-1) \,, \quad 2 \leq x \leq L-1 \,, \\
0 &=& (6-E(k)) f(1) - 2 f(2) \,, \quad \quad
0 = (6 - E(k)) f(L) - 2 f(L-1) \,.
\nonumber
\end{eqnarray}
Using the first relation we can deduce
\begin{eqnarray}
\label{EV}
E(k) = 4(1- \cos k) \,.
\end{eqnarray}
We may then solve for $A(k)$ and $\tilde{A}(k)$ as follows.  All
the relations (\ref{VRelations}) are consistent with the
single result (\ref{EV}) only if the exceptional second and third
equations in (\ref{VRelations}) take the same form as the first,
with auxiliary quantities $f(0)$, $f(L+1)$ introduced.  This
follows only if
\begin{eqnarray}
f(0) = - f(1) \,, \quad \quad f(L+1) = - f(L) \,.
\end{eqnarray}
If we plug this into the Bethe ansatz (\ref{Bethef}), these equations imply
\begin{eqnarray}
A(k) (1 + e^{ik}) &=& - \tilde{A}(k) (1 - e^{-ik}) \,, \\
A(k) e^{ikL} (1 + e^{ik}) &=& - \tilde{A}(k) e^{-ikL} (1 + e^{-ik}) \,,
\end{eqnarray}
which can be seen to have a solution only if
\begin{eqnarray}
\label{VBE}
e^{2ikL} = 1 \,.
\end{eqnarray}
This is the Bethe ansatz equation for this impurity.  We see that
in fact the boundary interactions are trivial and an elementary
quantization of the quasi-momentum $k$ is implied:
\begin{eqnarray}
{\cal B}_1 = {\cal B}_2 = 1 \,, \quad \quad k = {\pi n \over L} \,.
\end{eqnarray}
Meanwhile one finds the solution $A(k)/\tilde{A}(k) = - (1 +
e^{-ik})/(1 + e^{ik})$.  The energy eigenstates (\ref{Bethef}) can
hence be written in terms of the integer $n$:
\begin{eqnarray}
|W(n)\rangle = a_n \sum_{x = 1}^L  \sin \left[ {\pi n \over L}
\left(x- 1/2 \right) \right]  |W(x) \rangle\,
\end{eqnarray}
up to some normalization $a_n$, with the associated anomalous
dimension
\begin{eqnarray}
\gamma_{W(n)} = {g^2 N \over 4 \pi^2} \left[ 1 - \cos \left({\pi n \over L } \right)  \right] \,.
\end{eqnarray}
We see that the boundary conditions imposed on the spin waves are
that they go to zero at the ``end points" a half-step past the
last links on the chain;  these are Dirichlet boundary conditions.

This is precisely what we expect from the dual string theory point
of view.  In the string dual, this operator corresponds to an open
string living on a D5-brane; and oscillations in the three
$SO(3)_V$ directions normal to the brane must obey Dirichlet
boundary conditions at the endpoints.  Moreover, the periodicity
(\ref{VBE}) with no ${\cal B}$ factors tells us that the left and
right movers on the chain can be recast as just left moving
excitations on a closed chain of length $2L$;  this is nothing but
the familiar ``doubling trick" of open string theory.    We now
turn to a discussion of the $X^3$ excitation, which should be
associated with Neumann boundary conditions.

\subsection{Single Impurity Sector---$X^3$ Fluctuation}

We now consider an $X^3$ impurity, which we will abbreviate
as just $X$.  This lowers the $j_H$ charge to $j_H = L$ without turning
on a $j_V$ charge.  As before we define states with a single
impurity
\begin{eqnarray}
| X(x) \rangle \equiv \bar{q}_1 Z Z \ldots X \ldots Z q_2 \,,
\end{eqnarray}
with $X$ replacing $Z$ at site $x$. However, these states are not
a closed set under action of the Hamiltonian.  We may also create
states with $j_H = L$, $j_V = 0$ by flipping the spin of one of
the $q$ fields on the end of the chain, but leaving all bulk
excitations as $Z$:
\begin{eqnarray}
\label{qStates}
|q_1 \rangle \equiv \bar{q}_1 Z Z \ldots Z q_1 \,, \quad \quad
|\bar{q}_2 \rangle \equiv \bar{q}_2 Z Z \ldots Z q_2 \,.
\end{eqnarray}
We assume the energy eigenstates to be of the form
\begin{eqnarray}
\label{Betheg}
|X(k) \rangle \equiv \sum_{x = 1}^L g(x) |X(x)\rangle + \alpha |q_1 \rangle + \beta |\bar{q}_2 \rangle \,, \quad \quad g(k) = B(k) e^{ikx} + \tilde{B}(k) e^{-ikx} \,.
\end{eqnarray}
The analysis then follows as before. Requiring $|X(k) \rangle$ to
be an energy eigenstate
\begin{eqnarray}
(H - E(k)) |X(k) \rangle = 0 \,
\end{eqnarray}
leads to the equations
\begin{eqnarray}
0 &=& (4- E) \beta + 2 g(1) \,, \\
0 &=& (4 -E) g(1) + 4 \beta - 2 g(2) \,, \\
0 &=& (4 -E) g(x) - 2 g(x-1) - 2 g(x+1) \,, \quad 2 \leq x \leq L-1 \,, \\
0 &=& (4-E) g(L) - 2 g(L-1) - 4 \alpha \,, \\
0 &=& (4-E) \alpha - 2 g(L) \,.
\end{eqnarray}
The energy  has the same form as in the previous subsection, as
can be deduced from the generic ($2 \leq x \leq L-1$) equation
\begin{eqnarray}
E(k) = 4(1- \cos k) \,.
\end{eqnarray}
Again we proceed by attempting to cast the other equations in this
same form.  For the second and fourth equations, this can be done
using the definitions
\begin{eqnarray}
g(0) \equiv - 2 \beta \,, \quad \quad g(L+1) \equiv 2 \alpha \,.
\end{eqnarray}
The first and last equations also share this form if we introduce the fictitious
\begin{eqnarray}
g(-1) \equiv g(1) \,, \quad \quad g(L+2) \equiv g(L) \,.
\end{eqnarray}
These relations imply for the ansatz (\ref{Betheg})
\begin{eqnarray}
B(k) (e^{-ik} - e^{ik}) &=& \tilde{B}(k) (e^{-ik} - e^{ik}) \,, \\
B(k) e^{ikL} (e^{2ik} - 1) &=& \tilde{B}(k) e^{-ikL} ( 1 - e^{-2ik}) \,.
\end{eqnarray}
The first obviously forces $B(k) = \tilde{B}(k)$.  The second is then consistent only if
\begin{eqnarray}
\label{XBA}
e^{2ik(L+1)} = 1\,, \quad \quad k = { \pi n \over L + 1} \,.
\end{eqnarray}
Once again the Bethe ansatz equation indicates that the boundary reflection coefficients are trivial: ${\cal B}_1 = {\cal B}_2 = 1$. The chain is effectively one
link longer here, because of the participation of the $q$ fields in
the spin wave. The energy eigenstates can be written
\begin{eqnarray}
|X(n) \rangle = b_n \sum_{x=0}^{L+1} \cos \left( {\pi n x \over L+1} \right) |X(x) \rangle \,,
\end{eqnarray}
where we have defined
\begin{eqnarray}
|X(0) \rangle \equiv - (1/2) |\bar{q}_2 \rangle \,, \quad \quad |X(L+1) \rangle \equiv (1/2) |q_1 \rangle \,,
\end{eqnarray}
and the anomalous dimension is
\begin{eqnarray}
\gamma_{W(n)} = {g^2 N \over 4 \pi^2} \left[ 1 - \cos \left({\pi n \over L + 1} \right)  \right] \,.
\end{eqnarray}
We notice immediately that for these excitations, the boundary
conditions are Neumann.  Since $X$ corresponds to a fluctuation in
a direction in which the brane is extended, this is again exactly what we
expect from the dual string point of view; the possibility of
flipping the spin on the $q$ fields plays a fundamental role in
allowing the motion of the endpoints.  Again the left and right
movers together behave as a single closed chain, this time of
length $2(L+1)$.

We conclude that the integrability of the system allows us to see
with particular clarity the emergence of open strings from the
defect local operators.  Oscillations of the string are nothing
but spin waves propagating on the chain, reflecting with boundary
conditions that are equivalent to trivial transmission onto a
``doubled" closed chain.

\subsection{Plane Wave Limit}

The plane wave limit of the gravity dual to the dCFT was studied
by Lee and Park~\cite{LP}.  They considered a Penrose limit where
the defining null geodesic wound around the $S^2 \subset S^5$ on
which the D5-brane is wrapped; the associated light-cone vacuum
$|0, p^+ \rangle$ is just our Bethe reference state, the chiral
primary $|0 \rangle_L$, with $(\mu p^+ \alpha')^2 = J^2/(g^2 N)$,
$J=L$.  They furthermore studied single-impurity excitations, and
it is interesting to match these to our results.

Lee and Park identified an open string excitation in a Dirichlet
direction having oscillator number $n$ with the operator
\begin{eqnarray}
\label{LPD}
a^{\dagger, D}_n |0, p^+ \rangle \quad \leftrightarrow \quad {1 \over \sqrt{J}} \sum_{0}^J {\sqrt{2} \sin \left( {\pi n l \over J} \right) \over N^{J/2+1}} \bar{q}_1 Z^l  X^3 Z^{J-l} q_2 \,,
\end{eqnarray}
while a Neumann direction excitation took the form\footnote{The authors of \cite{LP} mention that operators analogous to (\ref{qStates}) with $q$ spins flipped could mix with (\ref{LPN}), but do not include them explicitly.}
\begin{eqnarray}
\label{LPN} a^{\dagger, N}_n |0, p^+ \rangle \quad \leftrightarrow
\quad {1 \over \sqrt{J}} \sum_{0}^J {\sqrt{2} \cos \left( {\pi n l
\over J} \right) \over N^{J/2+1}} \bar{q}_1 Z^l  X^3 Z^{J-l} q_2
\,.
\end{eqnarray}
They identified the anomalous dimension for both cases as
\begin{eqnarray}
\label{LPgamma}
\gamma_{plane} = {\pi g_s N n^2 \over 2 J^2} \,.
\end{eqnarray}
Using the relation $4 \pi g_s = g^2$, we find that the large-$L$
limit of our results indeed agrees with the conclusions (\ref{LPD})-(\ref{LPgamma}) of \cite{LP} for
the plane wave limit.

\section{Conclusions and Open Questions}
\label{ConclusionSec}

The presence of integrability inside four dimensional quantum field
theories is unexpected and fascinating.  We have seen that the sector
of open chain operators constructed from scalar fields in the
superconformal defect deformation of ${\cal N}=4$ Super Yang-Mills
possesses an integrable structure, taking the form of integrable
boundary conditions added to the structure of ${\cal N}=4$ SYM alone.
It is natural to hypothesize that this integrability holds for all
operators in the dCFT.  Given that theories closely related to ${\cal
N}=4$ are also integrable, one may speculate on how broad a class of
theories possesses an integrable structure, and to what extent
integrability is a useful concept for understanding other,
non-conformal gauge theories, for example those that confine.

The existence of the open chains also implies that in the gravity dual,
open string boundary conditions on D-branes can be integrable.  It is
natural to suspect that an $OSp(4|4)$ Yangian will survive in the open
Green-Schwarz worldsheet theory.  It would be interesting to verify the
presence of this symmetry, and to look for other D-brane configurations
that preserve integrability.

Finally, we have seen in the interactions of the spin waves in the
Bethe ansatz how properties of open strings like Dirichlet or Neumann
boundary conditions and the doubling trick arise in the field theory
dual as an open string is assembled from ``bits" of quantum fields on
the spin chain.  In this sense integrability, like supersymmetry, provides a crutch to
help us see the workings of duality.  It would be interesting to use
the Bethe ansatz to diagonalize more complicated states so as to learn
more about mesonic-type operators and their properties under
duality, and in greater generality to learn more about how strings are
assembled from field theory constituents.

\section*{Acknowledgments}

We benefited from discussions with David Berenstein, Neil Constable, Chris
Herzog, John McGreevy, Joe Polchinski, Radu Roiban, Marcus Spradlin, Herman
Verlinde, Anastasia Volovich, and Johannes Walcher.   This material is based upon
work supported by the National Science  Foundation under grants Nos.~0243680 (O.D.)  and PHY00-98395 (N.M.).  Any opinions, findings, and conclusions or recommendations expressed in  this material are those of the authors and do not necessarily
reflect the views of the National Science Foundation. The work of
N.M.\ was also supported by a National Defense Science and
Engineering Graduate Fellowship.

\section*{Appendix: Field Theory Conventions}

Our index conventions are as follows.  $\mu, \nu$ and $k, l$ are
4D and 3D Lorentz indices.  We denote the ${\bf 6}$
of $SO(6)_R$ with $i, j$; it decomposes under $SO(3)_H \times
SO(3)_V$ as ${\bf 6} \rightarrow ({\bf 3}, {\bf 1}) \oplus ({\bf
1}, {\bf 3})$, labeled with $I, J$ and $A, B$, respectively.
The ${\bf 4}$ of $SO(6)_R$ is $\alpha, \beta$, and it decomposes
as ${\bf 4} \rightarrow ({\bf 2}, {\bf 2})$; doublets of $SO(3)_H$
are denoted $m, n$ and doublets of $SO(3)_V$ are $a, b$.

We work in Euclidean space rotated from mostly-minus signature, where the $\gamma$-matrix conventions are those of~\cite{DFO}, and are such that Majorana spinors are real.

We write our adjoint fields in matrix notation, and color indices
are implicit throughout.  The action for ${\cal N}=4$ SYM is
\begin{eqnarray}
S_4 = \tf{1}{g^2} \int d^4x \,{\rm Tr}\, ( \tf12 F_{\mu\nu} F^{\mu\nu} - i \bar\lambda^\alpha \gamma^\mu D_\mu \lambda^\alpha +  D_\mu X^a D^\mu X^a  - \tf12 [X^a, X^b] + C_{\alpha\beta}^a \bar\lambda^\alpha [X^a, \lambda^\beta] ) \,,
\end{eqnarray}
with the covariant derivative $D_\mu X^a = \partial_\mu X^a - i
[A_\mu, X^a]$ and similarly for $\lambda^\alpha$.

The Yukawa tensor $C_{\alpha\beta}^a$ in the ${\bf 4} \times {\bf
4} \rightarrow {\bf 6}$ can be written explicitly as follows.
Split the fermions as $\chi^A \equiv \lambda^\alpha$, $A,\alpha =
1,2,3$, and $\lambda \equiv \lambda^4$, and assemble complex
scalars as $\Phi^A \equiv (X_V^A + i X_H^A)\sqrt{2}$ with
$X_V^{1,2,3} \equiv X^{1,2,3}$ and $X_H^{1,2,3} \equiv X^{4,5,6}$.
Then
\begin{eqnarray}
\int d^4x \, {\rm Tr} \, C_{\alpha \beta}^a \bar\lambda^\alpha [X^a, \lambda^\beta] &=& \\
i 2 \sqrt{2} \int d^4x \, {\rm Tr} \, (\bar\lambda [L \chi^A, \bar{\Phi}^A] - \bar\chi^A [R \lambda, \Phi^A] &-& \tf12 \epsilon_{ABC} \bar\chi^A [(L \Phi^C + R \bar{\Phi}^C), \chi^B] ) \,,
\nonumber
\end{eqnarray}
with $L \equiv (1 + \gamma^5)/2$, $R \equiv (1 - \gamma^5)/2$.

The $\lambda^\alpha$ are arranged into $\lambda_{am}$ transforming
in the $({\bf 2},{\bf 2})$ of $SO(3)_H \times SO(3)_V$:
\begin{eqnarray}
\lambda_{im} \equiv \lambda \delta_{im} - i \chi^A \sigma^A_{im} \,.
\end{eqnarray}
We use $z_\mu$ to indicate 4D coordinates and decompose them into
3D coordinates $y_k$ and transverse $x \equiv x_3$.  The massless
4D and 3D scalar propagators in Feynman gauge are
\begin{eqnarray}
\nonumber
\langle X(z_1) \, X(z_2) \rangle = {1 \over 2} \Delta_{12} \equiv {1 \over 2} {1 \over 4 \pi^2} { 1 \over (z_{12})^2} \,,  &&  \langle A_\mu(z_1) \, A_\nu(z_2) \rangle = {1 \over 2} \delta_{\mu\nu} \Delta_{12} \,, \\
\label{FermiProp}
\langle \lambda(z_1) \, \bar{\lambda}(z_2) \rangle \equiv {1 \over 2} S_{12} = {1 \over 2} \gamma^\mu\partial_\mu \Delta_{12}  \,,  && \langle q(y_1) \, \bar{q}(y_2) \rangle = {\cal D}_{12} \equiv   {1 \over 4 \pi} {1  \over |y_{12}|} \,,
\end{eqnarray}
where we have suppressed the color delta functions; the factors of
$1/2$ arise for the 4D fields because of the matrix notation we
are using. Since $\lambda$ is Majorana, we also have non-vanishing
$\langle \lambda (z_1) \, \lambda (z_2) \rangle$ and $\langle
\bar\lambda(z_1) \, \bar\lambda(z_2) \rangle$, determined by
$\bar\lambda = \lambda^T \gamma^0$.

We write the 3D fermion $\Psi$ in a 4D notation, as a four-component spinor satisfying
\begin{eqnarray}
P_+ \Psi = \Psi \,, \quad \quad P_- \Psi = 0 \,, \quad \quad P_\pm \equiv \tf12 \left( 1 \pm \gamma^5 \gamma^3 \right) \,,
\end{eqnarray}
where $\gamma^5 \equiv -i \gamma^0 \gamma^1 \gamma^2 \gamma^3$ is the chirality matrix.
The propagator for $\Psi$ is then
\begin{eqnarray}
\label{PsiProp} \langle \Psi(z_1) \bar\Psi(z_2) \rangle =
(\gamma^k P_+)  \,\partial_k {\cal D}_{12} \equiv \hat{s}_{12} \,.
\end{eqnarray}
The projection matrices $P_\pm$ obey
\begin{eqnarray}
[P_\pm , \gamma^k ] = 0 \,, \quad  \gamma^3 P_\pm = P_\mp \gamma^x \,, \quad
P_+^2 = P_-^2 = 1 \,, \quad P_+ P_- = 0 \,.
\end{eqnarray}

Propagators for $\lambda_{im}$ can be written in terms of (\ref{FermiProp}) as
\begin{eqnarray}
\nonumber
\langle \lambda_{im}(z_1) \, \bar\lambda_{nj}(z_2) \rangle &=&  2 \langle
\lambda(z_1) \, \bar\lambda(z_2) \rangle \delta_{ij} \delta_{mn} \,, \\
\label{FlavorProp}
\langle \lambda_{im}(z_1) \, \lambda_{jn}(z_2) \rangle &=&  2 \langle
\lambda(z_1) \, \lambda(z_2) \rangle \epsilon_{ij} \epsilon_{mn} \,, \\
\langle \bar\lambda_{mi}(z_1) \, \bar\lambda_{nj}(z_2) \rangle &=&  2 \langle
\bar\lambda(z_1) \, \bar\lambda(z_2) \rangle \epsilon_{ij} \epsilon_{mn} \,.
\nonumber
\end{eqnarray}
In differential regularization we use the replacements (neglecting quadratic divergences):
\begin{eqnarray}
{1 \over z^4} &\rightarrow& - {1 \over 4} \square   { \log z^2 M^2 \over z^2} \,, \\
{1 \over z^6} &\rightarrow& - {1 \over 32} \square \square { \log z^2 M^2 \over z^2} \,, \\
{1 \over |y|^3} &\rightarrow& - \nabla^2 { \log M|y| \over |y|} \,, \\
{1 \over |y|^5} &\rightarrow& - {1 \over 6} \nabla^2 \nabla^2 { \log M|y| \over |y|} \,,
\end{eqnarray}
where $\square$ and $\nabla^2$ are the 4D and 3D Laplacians, respectively.

\end{document}